\documentstyle[special,12pt]{article}
\begin{document}

\heading{The Palomar Digital Sky Survey (DPOSS) \footnotemark} 

\footnotetext{
To appear in {\sl Wide Field Surveys in Cosmology}, proc. XIV IAP Colloq.,
eds. S. Colombi and Y. Mellier, in press.
}

\par\medskip\noindent

\author{S. G. Djorgovski$^1$,
~R. R. Gal$^1$,
~S. C. Odewahn$^1$,
~R. R. de Carvalho$^2$,
~R.Brunner$^1$,
~G. Longo$^3$,
~R. Scaramella$^4$
}

\address{Palomar Observatory, Caltech, Pasadena, CA 91125, USA}
\address{Observatorio Nacional, CNPq, 20921 Rio de Janeiro, Brasil}
\address{Osservatorio Astronomico di Capodimonte, I-80131 Napoli, Italy}
\address{Osservatorio Astronomico di Roma, I-00040 Monteporzio, Italy}

\begin{abstract}
We describe DPOSS, a new digital survey of the northern sky, based on the
POSS-II photographic sky atlas.  The survey covers the entire sky north of
$\delta = -3^\circ$ in 3 bands, calibrated to the Gunn $gri$ system, reaching
to equivalent limiting magnitude of $B_{lim} \sim 22^m$.  As a result of
the state-of-the-art digitisation of the plates, detailed processing of the
scans, and a very extensive CCD calibration program, the data quality exceedes
that of the previous photographically-based efforts.  The end product of the
survey will be the Palomar-Norris Sky Catalog, anticipated to contain $> 50$
million galaxies and $> 2$ billion stars, down to the survey classification
limit, $\sim 1^m$ above the flux detection limit.  Numerous scientific projects
utilising these data have been started, and we describe briefly some of them;
they illustrate the scientific potential of the data, and serve as the
scientific verification tests of the survey.  Finally, we discuss some 
general issues posed by the advent of multi-terabyte data sets in astronomy.
\end{abstract}

\section{ Introduction }

The Palomar Digital Sky Survey (DPOSS) represents a digital version of the
POSS-II photographic sky atlas \cite{Reid}.
It is based on the scans of the original plates, done at the Space Telescope
Science Institute \cite{Lasker}, \cite{McLean}.
The final result of this effort will be a catalog of all objects detected down
to the survey limit, the Palomar-Norris Sky Catalog (PNSC).
For more details, see, e.g., \cite{Djorg98}.

The goal of this project is to provide a modern, uniform digital data set
covering the entire northern sky in 3 survey bands (photographic $JFN$,
calibrated to Gunn $gri$), with a photometric and object-classification
accuracy of sufficient quality to enable a wide range of scientific
follow-up studies.  

The survey processing and calibration were designed to extract all of the
information present in the plates. 
Our tests indicate that the resulting DPOSS and PNSC data are superior to
other, photographically based sky surveys in the past in terms of the
photometric quality, uniformity, depth, and object classification accuracy.  
We believe that this improvement is due to a combination of factors:
(a) a superior scanning process, which minimizes scattered light problems,
while maintaining the full angular resolution and dynamical range present in
the plate data;
(b) an unprecedented amount of CCD calibrations and object classification
training data sets;
(c) detailed processing which extracts all of the information present in
the plate scans, including various advanced techniques for object
classification. 

While the future fully digital sky surveys (e.g., the Sloan DSS; \cite{Gunn}) 
will provide photometric data of superior quality and depth than what is
possible with photographic technology, such data will not be generally
available over a substantial area on the sky for some years to come.  In the
meantime, DPOSS and PNSC can provide to the astronomical community data of 
adequate quality for many scientific studies, e.g., as envisioned for 
the SDSS.  Moreover, DPOSS will also cover the lower Galactic
latitudes, where SDSS data will not exist. 

We have already started a number of scientific follow-up studies using DPOSS
data, some of which are described below.  We find that in order to create
scientifically useful data, it is absolutely essential to try to use the data
for some actual scientific studies right away, and thus discover any possible
problems which may be affecting the data.  No other kind of a priori
constructed tests can really provide the necessary feedback.  Scientifically
viable catalogs cannot be constructed without such validation tests.  Yet, this
is a lesson which is not always appreciated by science managers or funding
agencies. 

In the process of analysing DPOSS and creating the PNSC, we have encountered
the problems posed by the handling, management and manipulation of
multi-Terabyte data sets, which are now appearing in astronomy and many other
fields.  Beyond the simple handling and maintenance of such data sets, a much
more interesting set of problems arises on how to explore or data-mine them
effectively -- how to convert this great abundance of information into actual
scientific results.  Traditional data processing techniques are simply
inadequate for these tasks.  We describe below some of our experiences and
ideas on how to approach these problems.

\section{ Survey Specifics }

The POSS-II covers the entire northern sky ($\delta > -3^\circ$) with 894
overlaping fields ($6.5^\circ$ square each with $5^\circ$ spacings), and,
unlike the old POSS-I, with no gaps in the coverage.  Approximately half
of the survey area is covered at least twice in each band, due to plate
overlaps.  Plates are taken in three bands:
blue-green, IIIa-J + GG395, $\lambda_{eff} \sim 480$ nm;
red, IIIa-F + RG610, $\lambda_{eff} \sim 650$ nm; and
very near-IR, IV-N + RG9, $\lambda_{eff} \sim 850$ nm.
Typical limiting magnitudes reached are $B_J \sim 22.5$, $R_F \sim 20.8$, and
$I_N \sim 19.5$, i.e., $\sim 1^m - 1.5^m$ deeper than the POSS-I.  The image
quality is improved relative to the POSS-I, and is comparable to the southern
photographic sky surveys.
As of the summer of 1998, the plate taking is 100\% complete in J, $\sim 97$\%
complete in F, and $\sim 83$\% complete in N, and it should be finished by the
spring of 1999. 

The original survey plates are digitized at STScI, using modified PDS scanners
\cite{Lasker}.
The scanning is closely following the plate taking in terms of the completeness.
The STScI group is using these data and an independent processing software to 
generate the second HST guide-star catalog, GSC-2.  The plates are scanned with
15-micron (1.0 arcsec) pixels, in rasters of 23,040 square, giving $\sim 1$
GB/plate, or $\sim 3$ TB of pixel data total for the entire digital survey
(DPOSS).  Preliminary astrometric solutions are good to r.m.s. $\sim 0.5$
arcsec, and should eventually reach r.m.s. $\sim 0.3$ arcsec.  An independent
digitisation of these plates is planned at USNOFS, by Monet {\it et al.} 

These data are superior to the widely available DSS scans, for several reasons:
(a) the original plates are scanned, rather than second copies; (b) the plates
have a finer grain, better image quality, and reach $\sim 1^m$ deeper than
POSS-I; (c) the scans have an improved dynamical range and finer pixels, viz.,
1.0 arcsec instead of 1.73 arcsec used in the DSS.  All of the cataloging work
at STScI and Caltech is done on the original scans, rather than the compressed
images available through the appropriate web servers and in the anticipated
future digital media distributions. 

Our ongoing effort at Caltech and the sites in Italy (OAC, OAR) and Brasil
(ON/CNPq) is to process, calibrate, and catalog the scans, with the detection
of all objects down to the survey limit (approximately to the equivalent
limiting magnitude of $B_J \sim 22^m$), and star/galaxy classifications
accurate to 90\% or better down to $\sim 1^m$ above the detection limit.

We use SKICAT, a novel software system developed for this purpose
\cite{WeirSKICAT}, \cite{WeirIAU}.  It incorporates some standard astronomical
image processing packages, commercial Sybase DBMS, as well as a number of
artificial intelligence (AI) and machine learning (ML) modules.  We measure
$\sim 60$ attributes per object on each plate, a subset of which is used for
object classifications. 

An essential part of this effort is the extensive CCD calibration program,
conducted mainly at the Palomar 60-inch telescope (with approximately 40 nights
per year allocated to this program), and some additional data on the equatorial
fields obtained at CTIO and ESO.  The data are calibrated in the Gunn $gri$
system; this will also make a tie-in with the future Sloan DSS easier.  There
is a good bandpass match (and thus small color terms) for the F and N plates
and the $r$ and $i$ bands, but the J plates extend considerably bluer than the
$g$ band.  These CCD data serve a dual purpose: for magnitude calibrations, and
as the training data sets for our automated object classifications, described
below.  Only the best-seeing data are used for the latter purpose.  The typical
limiting magnitudes of calibrated plate data are
$g_{lim} \approx 21.5^m$, 
$r_{lim} \approx 20.5^m$, and
$i_{lim} \approx 19.8^m$ 
(the CCD images reach $\sim 1^m - 2^m$ deeper).

We obtain a median of 2 CCD fields per plate, in all 3 bands.  This is an
unprecedented amount of CCD calibration for a photographically based sky
survey.  Yet, it is a minimum necessary in order to achieve the photometric
uniformity of $\sim 5 - 10$\% in magnitude zero-points, both across the
individual plates, and between the plates, as demonstrated in our early tests
\cite{WeirGALCO}.  We also use spatially filtered measurements of the local sky
intensity to determine the ``flatfield'' corrections for the plates; these are
partly due to the vignetting of the telescope optics, and partly to the
individual plate sensitivity variations.  Many projects, e.g., studies of the
large-scale structure, require magnitude accuracy and uniformity at this
level of accuracy or better.

Particular attention was paid to star-galaxy classification.  SKICAT uses
artificial induction decision tree techniques \cite{WeirCLASS}.  We are now
introducing additional neural-net (NN) based classifications \cite{OdewahnAJ},
\cite{OdewahnNN}. By using these methods and superior CCD data to train the AI
object classifiers, we are able to achieve classification accuracy of 90\% or
better (aiming for $\sim 95$\% or better) down to $\sim 1^m$ above the plate
detection limit; traditional techniques achieve comparable accuracy typically
only $\sim 2^m$ above the detection limit.  This effectively triples the number
of usable objects for most scientific applications of these data, since in most
cases one wants only either stellar objects or galaxies. 

Moreover, since the surface density of stars varies greatly over the sky, while
the surface density of galaxies (corrected for the extinction) remains roughly
constant, the contamination signal (e.g., stars misclassified as galaxies) can
vary greatly across the sky.  This is a major, but solvable problem for many
projects, and for {\it any} sky survey, yet it is seldom addressed with the
care it requires.  It is essential to calibrate and understand well the
statistical accuracy of object classifications in order to draw meaningful
conclusions from ostensible ``star'' or ``galaxy'' catalogs. 

Classification problems are present at all magnitudes, limited by the S/N at 
the faint end, by saturation at the bright end, and by crowding at
any flux level.  We are thus investigating the use of complementary object
classifiers optimized for different signal levels, optimal combining of object
parameters measured in different bandpasses, etc.

The final result of this effort will be the Palomar-Norris Sky Catalog (PNSC),
which will contain all objects down to the survey limit (equivalent to $B_{lim}
\sim 22^m$), with classifications and their statistically estimated accuracy
available down to $\sim 1^m$ above the detection threshold.  The catalog will
be confusion limited at low Galactic latitudes, where the surface density of
sources exceedes $\sim 20$ million per plate.  We estimate that the catalog
will contain $> 50$ million galaxies, and $> 2$ billion stars, including $\sim
10^5$ quasars.  The expected median redshift for the galaxies is $z \sim 0.2$,
reaching out to $z \sim 0.5$.

The catalog and its derived data products will be published electronically as
soon as the validation tests are complete, and our funding allows it, probably
starting in the early 1999. 
We note that the size of the DPOSS data set, in terms of the bits, numbers of
sources, and resolution elements, is $\approx 1,000 ~\times$ the entire IRAS
data set, and is $\approx 0.1 ~\times$ the anticipated SDSS data set.

\section{ Some Initial Scientific Applications }

This large new data set can fuel numerous scientific studies in the years to
come.  While the survey is not very deep by modern large-telescope
standards, it does cover $2\pi$ sterad, and it does so reasonably uniformly,
with a good wavelength baseline.  This enables several types of investigations:

\begin{itemize}
\item Optical identifications of sources detected in other wavelengths, e.g.,
in surveys ranging from radio through IR to x-ray, and the resulting
statistical multi-wavelength studies.  This could also lead to detections of
astrophysically interesting objects with extreme flux ratios (e.g., brown
dwarfs, ultraluminous IR galaxies, etc.). 

\item Statistical studies, such as the measurements of the large-scale
structure or Galactic structure, where the large numbers of sources can enable
meaningful fits of models with a large number of parameters and with small
Poissonian error-bars.  The essential requirement here is that the survey
calibrations are uniform and well understood, both in the terms of flux
calibration, and object classification. 

\item Searches for rare, or even previously unknown types of objects,
as defined by clustering in the parameter space, e.g., objects with unusual
colors and/or morphological structure, etc.  This may be the most intriguing
type of application, as it can lead to some really novel discoveries. 
\end{itemize}

We have already started a number of scientific projects along these lines
using DPOSS data.  They represent both our scientific motivation for doing
the work, and also serve as scientific verification tests of the data,
helping us catch and correct processing errors and improve and control the 
data quality.

Galaxy counts and colors in 3 bands from DPOSS can serve as a baseline for
deeper galaxy counts and a consistency check for galaxy evolution models. 
Our initial results \cite{WeirGALCO} show a good agreement with simple models
of weak galaxy evolution \cite{KGB} at low redshifts, $z \sim 0.1 - 0.3$.  We
are now expanding this work to a much larger area, to average over the local
large-scale structure variations.  We are also planning to cross-correlate our
galaxy counts and colors with the new Galactic extinction map \cite{SFD}; this
should lead to both an improved DPOSS galaxy catalog, and a better extinction
map. 

Another anticipated data product is a catalog of $\sim 5 \times 10^5$ brightest
galaxies on the northern sky (down to $B \sim 17^m$), with automated, objective
morphological classifications and at least some surface photometry information.
We have also started automated searches for low surface brightness galaxies
(in collaboration with J. Schombert), and for very compact, high surface
brightness galaxies, i.e., probing regions of the parameter space where the
selection effects tend to hide galaxies.  Follow-up redshift surveys of these
objects are now uder way at Arecibo and Palomar.

Our galaxy catalogs have been used as input for redshift surveys down to $\sim
21^m$, e.g., in the Palomar-Norris survey \cite{Small}, and several other
groups plan to use our catalogs for their own redshift surveys. 

Our preliminary investigation \cite{Brain} of the galaxy two-point correlation
function in 10 fields near the north Galactic pole, where the corrections due
to extinction and object misclassifications are minimal, indicate less power at
large scales than was found in the APM survey \cite{Mad}.  We suspect that
the differences may be due in part to an order of magnitude difference in the
amount of CCD calibrations used in the two surveys, and to the differences in 
star-galaxy classification accuracy.  Just about any instrumental effect or
calibration nonuniformity would add spurious power at large angular
scales.  This is a very important test for the scenarios of large scale
structure formation, and we want to be sure that we understand our data
fully before drawing any far-reaching cosmological conclusions.  The fact
that we have data in 3 bands in DPOSS should be of great utility here.

We are using DPOSS data to create an objectively and statistically well defined
catalog of rich clusters of galaxies \cite{Gal}. 
There are many cosmological uses for such clusters, and while the subjective
nature of the Abell catalog has been widely recognized as its major limitation,
many far-reaching cosmological conclusions have been drawn from it.  There is
thus a real need to generate well-defined, objective catalogs of galaxy
clusters and groups, with well understood selection criteria and completeness. 

Our approach is to use color selection of galaxies which are more likely to
define denser environments, followed with the adaptive kernel smoothing of
the resulting surface density map.  Statistically significant peaks are then
found using a bootstrap technique.  We typically find $\sim 1 - 1.5$ cluster
candidates per $deg^2$, and recover all of the known Abell clusters in the
same fields.  Our cluster sample reaches about a factor of 2 deeper than
the Abell sample, and extends to lower richness, thanks to the greater depth
of the plates and the automated selection of overdense regions.  We are
also completing a redshift survey of about 100 of the newly selected clusters,
in order to independently quantify our selection and completeness criteria.

We estimate that eventually we will have a catalog of as many as 20,000 rich
clusters of galaxies at high Galactic latitudes in the northern sky, with a
median redshift $\langle z \rangle \sim 0.2$, and perhaps reaching as high as
$z \sim 0.5$.  We plan to use this cluster catalog for a number of follow-up
studies, including cluster clustering, cross-identifications with x-ray
selected samples, etc.  We are also conducting detailed studies of the known
galaxy clusters, e.g., their galaxy luminosity functions, morphology, etc.

Another ongoing project is a survey for luminous quasars at $z > 4$.  Quasars
at $z > 4$ are valuable probes of the early universe, galaxy formation, and the
physics and evolution of the intergalactic medium at large redshifts.  The
continuum drop across the Ly$\alpha$ line gives these objects a distinctive
color signature:  extremely red in $(g-r)$, yet blue in $(r-i)$, thus standing
away from the stellar sequence in the color space.  Traditionally, the major
contaminants in this type of work are red galaxies.  Our superior star-galaxy
classification leads to a manageable number of color-selected candidates, and
an efficient spectroscopic follow-up.  As of mid-1998, over 40 new $z > 4$
quasars have been discovered.  We make them available to other astronomers for
their studies as soon as the data are reduced.

Our initial results \cite{Ken95a}, \cite{Ken95b}, are the best estimates to
date of the bright end of the quasar luminosity function (QLF) at $z > 4$, and
are in excellent agreement with the fainter QLF evaluated in a completely
independent survey \cite{SSG}.  We confirm the decline in the comoving number
density of bright quasars at $z > 4$.  We find intriguing hints of possible 
primordial large-scale structure as marked by these quasars \cite{DjorgMOR},
but more data and tests are needed to check this result.  Our follow-up
projects include a search for protogalaxies and possible protoclusters in
these quasar fields, a new survey for high-redshift DLA absorbers, etc.

We have also started optical identifications of thousands of radio sources,
e.g., the VLA FIRST sources \cite{Becker}.  Our preliminary results indicate
that there are $\sim 400$ compact radio source ID's per DPOSS field, and we
expect a comparable number of resolved source ID's.  Eventually, we expect to
have $> 10^5$ ID's for the VLA FIRST sources, plus many more from other
surveys.  Our primary goal is to select radio-loud quasars at $z > 4$; to date,
2 such objects have been found \cite{Stern}.  
We have also obtained DPOSS IDs for a sample of several hundred flat-spectrum
radio sources, selected from the GB, TEX, and PKS samples.  For sources with
the spectral index $\alpha > -0.3$, which should be mostly quasars, the DPOSS
identification rate approaches 97\%.  We thus find that at most a few percent
of such radio sources may be completely obscured by dust \cite{Kollmeier}, 
in contrast to some other claims \cite{Webster}. 

In the area of statistical gravitational lensing studies, we have explored the
possibility of microlensing of quasars, by looking for a possible excess of
foreground galaxies near lines of sight to apparently bright, high-$z$ QSOs
from flux-limited samples.  We find at most a modest excess.  We are also 
planning to use our galaxy counts to explore the possible lensing magnification
of background AGN by foreground large scale structure \cite{BartSch}. 

Much remains to be done in the area of Galactic astronomy.  Star counts as a
function of magnitude, color, position, and eventually proper motion as well,
fitted over the entire northern sky at once, would provide unprecedented
discrimination between different Galactic structure models, and constraints on
their parameters.  With $\sim 2 \times 10^9$ stars, such studies would present
a major advance over similar efforts done in the past.  With the inclusion of
IR data such studies can be made much more powerful.  A combination of optical
data from DPOSS and IR data from 2MASS can be very efficient in a search for
brown dwarfs, or stars with unusual colors or variability. 

We are now applying the same techniques we use to search for galaxy clusters
to our star catalogs, in an objective and automated search
for sparse globulars in the Galactic halo, tidal disruption tails of former
clusters, and possibly even new dwarf spheroidals in the Local Group (recall
the Sextans dwarf, found using similar data \cite{Irwin}).  We are also using
DPOSS data to map the tidal cutoff regions of selected Galactic globulars
\cite{Zaggia}.

This is just a modest sampler of the scientific uses of DPOSS, which are 
already under way.  We can expect much more in the years to come.

\section{ Technology Issues: ~How to Harvest the Abundance 
of Multi-Terabyte Astronomical Data Sets }

The advent of multi-terabyte astronomical data sets will change profoundly
the face of astronomy.  We are facing not only terabytes of raw data (or
pixels), but terabytes of {\it reduced} data (or catalogs): archives containing
$\sim 10^9 - 10^{10}$ objects with $\geq 10^2$ measured parameters each.  
At this time, several large digital sky surveys are planning to produce data
sets or archives of this size, but such data volumes will become more common
or even standard for individual projects or experiments.  This new wealth of
information will enable: 

\noindent{\bf 1. New astronomy:}  Doing statistical astronomy ``right'', not
limited by poissonian noise or data poverty; searching for rare or new
types of astronomical objects or phenomena; asking new kinds of questions as 
we characterize the sky as a whole, rather than work with small samples of 
objects; and so on.  Combining giga-object surveys from different wavelengths
should be especially valuable.  This is a quantitatively and qualitatively
different enterprise, with a different set of requirements and goals from
the existing (and very useful) astronomical data centers and web tools like
Simbad, NED, SkyView, etc., which are mainly intended to provide data services
for individual objects or small samples of objects or fields.

\noindent{\bf 2. A new style of doing observational astronomy:}  These vast
data sets can sustain extended data-mining by numerous users, making uses of
the data which were not even conceived by the original data producers.  Anyone,
anywhere, with a computer and an internet connection will be able to do some
first-rate observational astronomy, without a need to access expensive or
exclusive telescopes or other facilities, without necessarily being associated
with an elite astronomical institution.  This will enable a broader sampling of
the community talents, change the sociology of astronomy, require new kinds
of technical skills, and change the way that astronomical research is done. 
In a way, there is a parallel with analytical theory and numerical simulations,
which ideally can go hand-in-hand, but can be (and usually are) practiced by
different scientists with different skills.  So we will have traditional
observers and data miners, and various species in between. 

The opportunities are great, but so are the technical challenges.  Traditional
astronomical data processing tools are completely inadequate for the tasks at
hand.  We, as a community, need to develop or adopt from elsewhere a whole
new set of tools and skills for an effective exploration of multi-terabyte
data sets.  In some sense, this is similar to the digital imaging revolution 
of $\sim 15 - 20$ years ago, when astronomy changed from dealing with small
1-dimensional data sets to dealing effectively with ever larger digital 
images or data cubes at any wavelength.  Eventually several major, standard
packages evolved, but this was not always a fully rational or optimal process. 
Back then we had to learn about things like PSF and isophote fitting, and now
we have to learn about querying and exploring large databases. 

These vast data sets will be changing constantly, as new or better calibrations
or reprocessing algorithms are introduced.  Different archives will be combined
and recombined, creating new data sets, permanent or transitory.  
The very concept of an astronomical catalog is changing, from a fixed set of
printed volumes, to a permanently evolving database which has to be accessed
and explored in some non-trivial manner.  The largest astronomical catalogs
in the past contained $O(10^5)$ objects with $O(10^1)$ parameters each, were
never recalibrated, and required several thick printed volumes.  The new
multi-terabyte catalogs will never be printed, but rather, will reside in
network-accessible archives, and would always have to come along with the
software tools necessary for their use and exploration.  Specifically, we need:

\noindent{\bf 1. New data structuring standards and formats.}  
This would
enable an easier interchange and matching of data sets, and commonality of
software tools.  In addition to the formulation of computing standards (e.g.,
relational vs. object-oriented databases, data exchange formats,
standard user interfaces and query languages, etc.), we should think about
new astronomical conventions, which would be optimised for our computing
needs, rather than be based on fossilized conventions from the past; a prime
example of this is the Hierarchical Triangulated Mesh for sky partitioning
\cite{HTM}.
The point is that we should structure the data in a way which would make the
most common types of queries work fast, and enable the new software tools to
sift through the data quickly and effectively.

\noindent
{\bf 2. New tools and expertise to navigate and explore these data spaces.}  
This would include fast ways to query the data in multidimensional parameter
spaces (with tens or hundreds of data dimensions), and the associated novel
data visualisation problems (including perhaps virtual reality walks through
the parameter spaces).  On a more sophisticated level, we need tools for an
automated clustering analysis and classification, both supervised (e.g., where
the program is trained on a set of examples of what to look for) and
unsupervised (where the program decides in some autonomous, statistically
justified fashion how many different kinds of objects are there in the data
space, and what they are).  We would probably want to make use of intelligent
software agents (a simple example are some of the better web search engines)
which would search through our data parameter space looking for desired kinds
of objects (e.g., clusters of galaxies defined in some objective manner), or
discrepant objects standing away from the bulk of the data points, and so on. 
We should make a good use of AI and ML techniques, and other modern data-mining
methods for true machine-assisted discovery. 

This long list of desiderata amounts to a new information infrastructure for
astronomy.  Virtual observatories are coming, and we should try to design them
right from the start, avoiding costly mistakes and unnecessary replication of
effort.  Many of the necessary tools already exist; similar problems are faced
by just about any other data-intensive discipline, and astronomers can benefit
greatly in this task from collaborations with computer scientists.  One thing 
is certain: we are entering a new era of information-rich astronomy, and we
should be ready for the abundance it has to offer.  More is {\it different}. 


\acknowledgements{
The DPOSS/PNSC cataloging effort at Caltech is supported by a generous grant
from the Norris Foundation.  Some of the software technology development has
been supported by the grants from NASA, JPL, and Caltech.  SGD also wishes to
acknowledge support from the Bressler Foundation.  We are indebted to the
entire POSS-II photographic survey team, the scanning team at STScI, to Palomar
Observatory for generous allocations of telescope time used for CCD
calibrations, and to our numerous collaborators and students whose work is
making DPOSS and PNSC become reality.  This work is a part of the CRONARio
collaboration.  Finally, SGD wishes to thank the
conference organizers for their hospitality, and to acknowledge 
the valuable guidance provided by \cite{Wells}.
}

\vfill
\end{document}